\documentclass[aps, pra, twocolumn,10pt,showkeys]{revtex4-1}
\usepackage[unicode]{hyperref}
\usepackage{dcolumn}
\usepackage{bm}
\usepackage{graphicx}
\usepackage{amsmath}
\usepackage[utf8x]{inputenc}
\usepackage[english]{babel}
\begin{document}

\title{Two-temperature momentum distribution in a thulium magneto-optical trap}

\author{E.\,Kalganova,$^{1}$ O.\,Prudnikov,$^{2,\,3}$ G.\,Vishnyakova,$^{1,\,4}$ A.\,Golovizin,$^{1,\,4}$ D.\,Tregubov,$^{1,\,4}$ D.\,Sukachev,$^{1,\,5}$ K.\,Khabarova,$^{1,\,4}$ V.\,Sorokin,$^{1,\,4}$ N.\,Kolachevsky$^{1,\,4}$}                     

\affiliation{
$^1$\,P.N.\,Lebedev Physical Institute, Leninsky prospekt 53, 119991 Moscow, Russia\\
$^2$\,Institute of Laser Physics, Siberian Branch, Russian Academy of Sciences, Prospekt Akademika Lavrentyeva 13/3, 630090 Novosibirsk, Russia\\
$^3$\,Novosibirsk State University,  Pirogova St. 2, 630090 Novosibirsk, Russia\\
$^4$\,Russian Quantum Center, Novaya St. 100A, Skolkovo, 143025 Moscow, Russia\\
$^5$\,Harvard University, Department of Physics, Oxford St. 17, Cambridge, 02138 Massachusetts, USA\\
}
\date{today}

\begin{abstract}
  Second-stage  laser cooling  of  thulium atoms at the 530.7\,nm transition with a natural linewidth of $350$~kHz offers an interesting possibility to study  different regimes of a magneto-optical trap (MOT). The intermediate value of the spectral linewidth of the cooling transition allows the observation of three distinct regimes depending on the intensity and detuning of the cooling beams, namely, the  ``bowl-shaped'' regime when  light pressure force competes with gravity, the ``double structure'' regime with interplay between the Doppler and polarization-gradient (sub-Doppler) cooling, and the  ``symmetric'' regime when Doppler cooling dominates over sub-Doppler cooling and gravity. The polarization-gradient cooling manifests itself  by a two-temperature momentum distribution of atoms  resulting in a  double-structure of the spatial  MOT profile  consisting of a cold central fraction  surrounded by a  hot halo. We studied the double structure regime at different saturation parameters and compared observations with calculations based on semiclassical and quantum approaches. The quantum treatment adequately reproduces experimental results if the MOT magnetic field is properly taken into account.
\end{abstract}

\maketitle
\section{Introduction}
\label{intro}
Since the first demonstration of laser cooling of Erbium atoms in 2006 \cite{Er_36MHz}, the interest in ultracold hollow-shell lanthanides is continuing to grow.
In addition to the relative simplicity of laser cooling and availability of laser sources, most of these atoms possess a large ground-state magnetic dipole moment which makes them promising candidates for studies of long-range anisotropic interactions \cite{Dipolar_interaction}.
 Recent advances in the preparation and control of laser-cooled ensembles of Dy \cite{Dy_2kHz+BEC}, Er \cite{Er_190kHz}, Tm \cite{Tm_second-stage_cooling, Tm_UFN_2016}, and Ho \cite{Ho} provide access to the physics of  dipolar quantum gases \cite{Dipolar_BEC,Dipolar_Fermi},  low-field Feshbach resonances \cite{Er_and_Dy_Feshbah} and quantum simulations of unexplored many-body phenomena \cite{Baier201}.
Inner-shell  transitions in lanthanides can also be used for optical frequency metrology \cite{Tm_clock_PRA, Er_clock}.

 Deep laser cooling of lanthanides  in a magneto-optical trap (MOT) is often done in two steps.
  For Zeeman slowing and  the first-stage MOT, a strong transition  lying in the blue spectral range around 400\,nm is typically used \cite{Er_36MHz, Dy_MOT, Tm_MOT, Ho}. That allows one to reach temperatures around 100\,$\mu$K.
Lower temperatures are  achieved using a second-stage MOT operating on a spectrally narrow or intermediate transition \cite{Sr_two-stage_cooling}. Two-stage laser cooling of the hollow-shell lanthanides was successfully demonstrated in Er \cite{Er_8kHz}, Dy \cite{Dy_2kHz+BEC} and Tm \cite{Tm_second-stage_cooling}. After the second stage cooling, atoms  are  loaded into an optical dipole trap or in an optical lattice for further studies.

For nearly all transitions involved  in laser cooling of hollow-shell lanthanides, the magnetic Land\'e $g$-factors of the upper and lower cooling levels are  close to each other \cite{Lanthanides_spectrum,Dy_spectrum}. As a result, efficient sub-Doppler cooling was observed directly in the first-stage MOT of Tm \cite{Tm_Sub-Doppler}, Dy \cite{Dy_double_structure}, Er \cite{Er_Sub_Doppler} and Ho \cite{Ho} even in the presence of a strong magnetic field gradient. However, sub-Doppler cooling in the second-stage MOT has not been  reported yet.

For the second-stage laser cooling of Tm atoms we use the transition at 530.7\,nm with a natural spectral linewidth of $\Gamma = 2 \pi \times 350$\,kHz [Fig. \ref{Levels}]. The intermediate transition linewidth and a small difference of magnetic Land\'e $g$-factors  allows us to observe competition between Doppler and sub-Doppler cooling as well as the interplay of gravity and light pressure force directly in the MOT. Depending on the saturation parameter  $S$, we can observe three distinct regimes: the regular ``symmetric'' regime dominated by Doppler cooling, the  ``bowl-shaped'' regime with the strong influence of gravity on MOT performance, and the most unusual  ``double structure'' regime where Doppler and sub-Doppler cooling mechanisms compete. To describe  the double structure regime we used semiclassical \cite{sub-Doppler_theory} and quantum \cite{Prudnikov2007,Prudnikov2011} simulations of the cooling process. The quantum approach takes into account recoil effects and, contrary to semiclassical approach,  adequately reproduces the momentum distribution in the cloud.

In this manuscript, we describe three MOT regimes  and present experimental data (Section \ref{regimes}). Section \ref{simulations} shows results of the semiclassical and quantum treatments and comparison with the experiment. Some conclusions are summarized in Section \ref{conclusion}.

\section{MOT regimes}
\label{regimes}

\begin{figure}[t]
\resizebox{0.45\textwidth}{!}{
\includegraphics{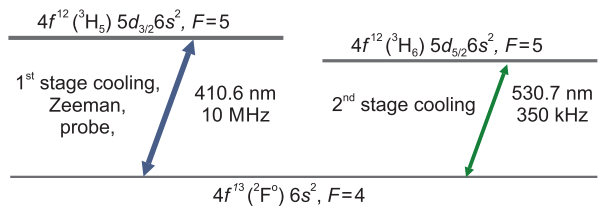}
}
\caption{Relevant energy levels of thulium. The strong transition
at 410.6\,nm is used for the first-stage laser cooling, the Zeeman slowing and the imaging, the weak transition at 530.7\,nm is used for the second-stage cooling.}

\label{Levels}
\end{figure}

We observe all three MOT regimes introduced in Section \ref{intro} by varying the  saturation parameter $S = S_0/(1+4\Delta^2/\Gamma^2)$. Here $S_0=I/I_\textrm{\rm sat}$ is  the saturation parameter on the exact resonance  with $I$ being the on-axis single-beam intensity and $I_\textrm{\rm sat}$ being the saturation intensity. The saturation intensity is defined as $I_\textrm{\rm sat} = 2 \pi^2 c \hbar \Gamma/ 3 \lambda^3$, where $\lambda$ is the wavelength of light, $c$ is the speed of light, $\hbar$ is the reduced Planck constant, and $\Gamma$ is the natural linewidth. For the second-stage Tm cooling transition $I_\textrm{\rm sat} = 0.32$\,mW$/$cm$^2$. The detuning of the laser frequency $\omega$ from the atomic resonance frequency $\omega_0$ is denoted as $\Delta = \omega -\omega_0$.

We use the conventional MOT configuration with three orthogonal pairs of circularly polarized cooling beams (the vertical beams  are aligned with gravity).  During the first cooling stage, atoms are trapped using  both blue ($410.6$\,nm) and green ($530.7$\,nm) cooling radiation simultaneously [Fig. \ref{Levels}]. Then radiation at 410.6\,nm (Zeeman and first-stage MOT cooling light) is blocked and atoms are further cooled only by 530.7\,nm light for  $\tau_{\rm sc}=80$\,ms  (the second-stage MOT). After the second-stage cooling and a period of ballistic expansion, the cloud is imaged on a CCD-camera using a short pulse of resonant 410.6~nm light.  More details about the experimental configuration are given in \cite{Tm_second-stage_cooling}.
To study different  MOT regimes, we perform several experiments with significantly different intensities and detunings of 530.7\,nm light.

We measure that the second-stage MOT reaches a steady state faster than in 40\,ms  for the whole range of parameters used in this work which agrees with the theoretical estimations based on ref. \cite{Doppler_theory}.
Accordingly, we choose the duration of the second-stage cooling ($\tau_{sc}$) to be longer than 40\,ms.
This assures that the first-stage cooling only determines number of atoms trapped in the second-stage MOT but does not affect its dynamics.

\begin{figure}[t]
\resizebox{0.45\textwidth}{!}{
\includegraphics{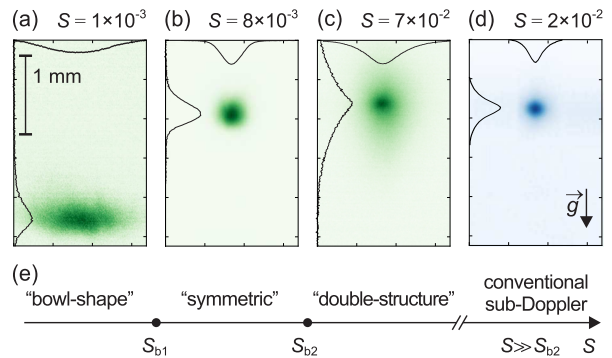}
}
\caption{(a-c) Images of the second-stage MOT cloud for (a) the bowl-shaped regime with  $\Delta = -7\Gamma,\, S_0 = 0.1$; (b) the symmetric regime, $\Delta = -2.4\Gamma,\, S_0 = 0.1$;  (c) the double structure regime, $\Delta = -11\Gamma,\, S_0 = 37$. (d) Image of the first-stage MOT cloud for the conventional sub-Doppler regime with $\Delta = -1\Gamma,\, S_0 = 0.1$.
The density profiles for  vertical and horizontal axes are shown;
$g$ denotes the direction of the gravity.  (e) MOT regimes depending on the saturation parameter value. The boundaries are $S_{\rm b1}=2\times10^{-3}$, $S_{\rm b2}=10^{-2}$ for the second-stage MOT (a-c) and  $S_{\rm b1}=5\times10^{-5}$, $S_{\rm b2}=7\times10^{-4}$ for the first-stage MOT (d).}

\label{Cloud_shape}
\end{figure}

Since the  magnetic field gradients are different for horizontal and vertical directions (the anti-Helmholtz coils are axially aligned along the vertical axis $z$), the vertical and horizontal momentum distributions  are also expected to be different. Henceforth, we will discuss  momentum distribution and the corresponding temperature only for the vertical coordinate $z$ (along gravity).
Momentum distribution and temperature for the horizontal coordinates qualitatively demonstrate the same behavior and further discussion is valid for the horizontal coordinates as well.

\begin{figure*}
\center{
\resizebox{0.95\textwidth}{!}{
\includegraphics{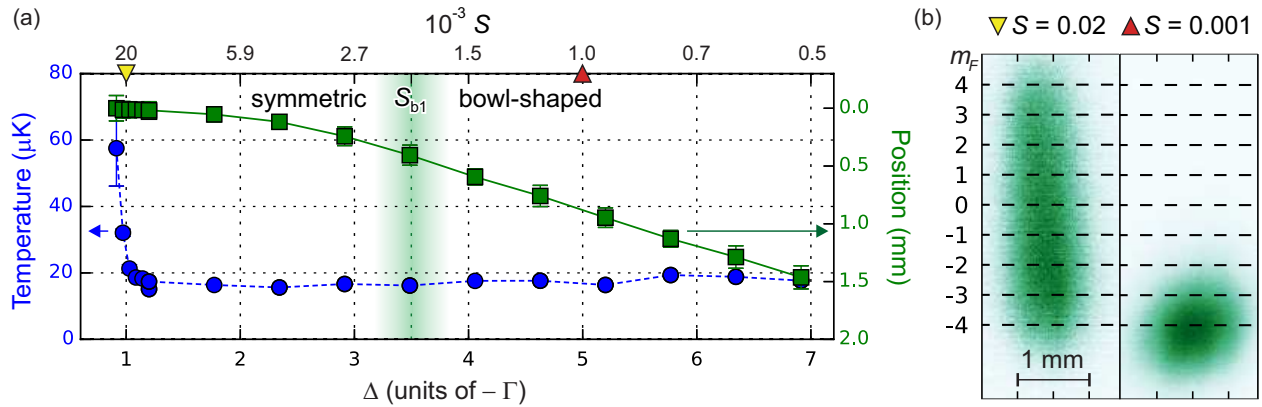}
}
\caption{a) The MOT temperature (blue circles) and the position (green squares) of the cloud center versus the laser frequency detuning $\Delta$ in units of the minus  natural linewidth $-\Gamma$. The resonant saturation parameter $S_0 = 0.1$. The respective values of detuning-dependent saturation parameters $S$ are given on the top axis. One can distinguish  the symmetric ($|\Delta| < 3.5\Gamma$) and the bowl-shaped (detuning $|\Delta| >  3.5\Gamma$) regimes. The green vertical boundary shows the case when the saturation parameter $S$ reaches  $S_{\rm b1}$. The lines are guidelines for the eye. b) The Stern-Gerlach experiment. The cloud image is taken after $6$\,ms of ballistic expansion in a magnetic field with a gradient of $0.4$\,T$/$m. All magnetic sub-levels $m_F$ are well populated in the  symmetric MOT regime (left, $S = 0.02$), while for the bowl-shape regime, strong optical pumping to the $m_F=-4$ sub-level is observed (right, $S = 0.001$).
Respective  saturation parameter values are marked by the yellow downward triangle and the red upward triangle on the panel (a).
}
\label{Stern-Gerlach}}
\end{figure*}

Unlike in the broad-line MOT, the intermediate and narrow-line MOT operation can be dramatically affected by gravitational force and the recoil effect.
The boundaries between three aforementioned regimes are roughly given by the following criteria for the saturation parameter $S$.
The first criteria is the balance between the  light pressure force and the gravitational force:
\begin{equation}
\frac{R S_0}{1+S_0+4(\Delta/\Gamma)^2}= 1,
\end{equation}
where $R = \hbar k \Gamma/2 m g$.
Here $k$ is the wave vector of the cooling light, $m$ is the atomic mass, and $g$ is the free fall acceleration.
Thus, the first boundary saturation parameter value is:
\begin{equation}
S_{\rm b1}=1/(R-1).
\end{equation}
In our case $R=500$ and the first boundary is $S_{\rm b1}=2\times10^{-3}$.

The second boundary corresponds to the case when the width of the  polarization-gradient force resonance in momentum domain \cite{sub-Doppler_theory}
\begin{equation}
\delta p = \frac{2m}{k}S \Gamma
\label{sub-D_width}
\end{equation}
becomes equal to the recoil momentum $\hbar k$:
\begin{equation}
S_{\rm b2}=\frac{\hbar k^2}{2m\Gamma}\,.
\end{equation}
Thus, the second boundary equals   $S_{\rm b2}=10^{-2}$.

The MOT regimes and corresponding atomic cloud density profiles recorded in our experiments are depicted in Fig. \ref{Cloud_shape}(a-c) and are described further in more detail.

\subsection{The bowl-shaped regime}
When $S<S_{\rm b1}$, the  light pressure force becomes comparable or smaller than the  gravitational force. The  cloud moves down to the region of nonzero magnetic field, reaching
the  equilibrium position  on the surface  where the Zeeman shift compensates for the light frequency detuning. At this surface, the light pressure force balances the gravitational force.

This causes three interesting effects.
 First, the cloud acquires a specific shape (the bowl-shape) as shown in Fig.\,\ref{Cloud_shape}(a). Second, the combined detuning becomes independent of the light frequency detuning $\Delta$. As a consequence, the cloud temperature also becomes independent of $\Delta$ [Fig.\,~\ref{Stern-Gerlach}(a)].  Third, atoms interact mostly with the upward propagating beam and become optically pumped to the lowest magnetic sub-level $m_F$, where $F$ is the total atomic angular momentum. We verify atomic spin-polarization using the Stern-Gerlach experiment [Fig.\,\ref{Stern-Gerlach}(b)], switching off the MOT fields and applying a vertical magnetic field gradient of about $0.4$\,T$/$m.
The optical pumping allows for the preparation of a spin-polarized atomic ensemble which is advantageous for further study of cold collisions.

This cooling regime was previously reported for Sr \cite{Sr_two-stage_cooling}, Dy \cite{Dy_bowl-shaped} and Er \cite{Er_8kHz}, and was analyzed in detail in \cite{Dy_bowl-shaped, Sr_narrow_line_cooling}.

\subsection{The symmetric regime}

If $S<S_{\rm b2}$, the width of the polarization-gradient resonance in the momentum domain [Eq. (\ref{sub-D_width})]
is small compared to the  recoil shift. As a result each photon scattering event pushes the atom out of the polarization-gradient resonance. In this case the  polarization-gradient cooling mechanism (sub-Doppler cooling) does not play a significant role and only Doppler cooling takes place.

The interval
$$
S_{\rm b1}<S<S_{\rm b2}
$$
corresponds to the MOT regime with the Doppler cooling mechanism playing the dominant role when the light pressure force is much stronger than the gravity. In this case, the atomic cloud has a symmetric elliptical shape [Fig.\,\ref{Cloud_shape}(b)] which we refer to as the ``symmetric regime''.

The symmetric ($S_{\rm b1}<S<S_{\rm b2}$) and the bowl-shaped ($S<S_{\rm b1}$) regimes were observed for the cooling light intensities of $0.02-0.1$\,mW$/$cm$^2$ per beam ($S_0 = 0.06-0.3$) and  detunings $\Delta$ in the range from $-1\Gamma$ to $-7\Gamma$.  Fig.\,\ref{Stern-Gerlach}(a) shows the dependence of the atomic cloud temperature $T$ on the frequency detuning $\Delta$ for $S_0=0.1$ and illustrates a transition between these two regimes.

\subsection{The double-structure regime}

If $S>S_{\rm b2}$, $\delta p$ becomes larger than the recoil momentum  $\hbar k$.
The polarization-gradient cooling starts impacting some fraction of an atomic ensemble, while the rest of the atomic cloud still possesses the  momentum distribution determined by the Doppler cooling. As a result, the momentum distribution of the atomic ensemble becomes significantly different from the  Maxwellian one, consisting  of sub-Doppler and Doppler fractions (see Section \ref{simulations}).
In the harmonic potential, this results in a two-component spatial density distribution [Fig.\,\ref{Cloud_shape}(c)], which can be adequately approximated by the sum of two Gaussian functions.

We observe such cloud shapes for intensities higher than 3 mW$/$cm$^2$ ($S_0 > 9$) and for frequency detunings  $\Delta$ from  $-4\Gamma$ to $-12\Gamma$.
The  ballistic expansion experiments show that the cloud consists of two atomic sub-ensembles possessing different temperatures.  There is a fraction of cold atoms which corresponds to a small central volume, and a hot fraction forming a spread halo. To determine temperatures of these two sub-ensembles, the cloud density profiles obtained during ballistic expansion are fitted by the two-Gaussian function with independent parameters. The fit gives us the temperature of the sub-Doppler fraction $T_{\rm SD}$, the temperature of the Doppler fraction $T_{\rm D}$ as well as the relative number  $\eta$ of the sub-Doppler cooled atoms.

\begin{figure}[!t]
\center{\resizebox{0.5\textwidth}{!}{
\includegraphics{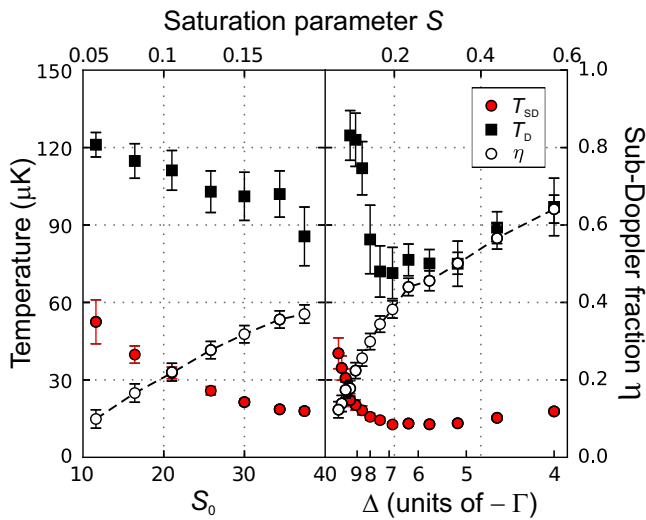}
}}
\caption{Temperature of the Doppler $T_D$ (black squares) and the sub-Doppler $T_{\rm SD}$(red circles) sub-ensembles depending on the light intensity and frequency detuning. Open circles show the fraction $\eta$ of the sub-Doppler cooled atoms. Left plot: intensity $I$ ($S_0 \propto  I$)  is changed at fixed $\Delta = -7\Gamma$. Right: light detuning $\Delta$ is changed at fixed  $I=11.8$~mW$/$cm$^2$ ($S_0=37$). In both experiments the  magnetic field gradient equals $b=0.07$ T$/$m. Top axes show the corresponding saturation parameter $S$.
}
\label{Experiment}
\end{figure}

We observe  the transition from the symmetric to the double structure regime  by changing the 530.7 nm light intensity $I$ ($S_0 \propto  I$) at fixed $\Delta$.  Similar behavior can be observed  by varying  $\Delta$ at constant $S_0$. The corresponding temperatures $T_{\rm D}$ and $T_{\rm SD}$  are shown in Fig.\,\ref{Experiment}.
 The fraction of the sub-Doppler cooled atoms $\eta$ (shown in the same plots) grows for larger saturation parameters $S$.
 One can interpret this as the result of  the  sub-Doppler resonance broadening [Eq. (\ref{sub-D_width})] which enhances the sub-Doppler cooling efficiency.

It is interesting to note that the temperature $T_{\rm SD}$ decreases with increasing light intensity;  the effect is clearly visible for saturation parameters $S$ in the range from 0.05 to 0.2. This is unusual for conventional pictures  of sub-Doppler \cite{sub-Doppler_theory} and Doppler cooling \cite{Doppler_theory}, where the increase of cooling intensity always results in the increase of the temperature of atoms. The observed behavior is interpreted in the next section using the framework of MOT quantum theory.

One should note that
MOT regimes  smoothly transform from one to another without clear boundaries. Nevertheless, theoretical estimates for $S_{\rm b1}$ and $S_{\rm b2}$ show reasonable agreement with our experimental  observations.

For comparison, let us  consider  Tm first-stage cooling at the spectrally broad transition  (410.6\,nm). The corresponding boundaries will be $S_{\rm b1}=5\times10^{-5}$  and $S_{\rm b2}=7\times10^{-4}$ (comparable to the values for a $^{87}$Rb  MOT \cite{Rb_data} of $S_{\rm b1}=9\times10^{-5}$ and $S_{\rm b2}=6\times10^{-4}$). For a typical first-stage Tm MOT \cite{Tm_Sub-Doppler}, we use  $S_0\approx 0.1$ and $\Delta\approx -\Gamma$, which  corresponds to the case $S=0.02\gg S_{\rm b2}$. In this case the width of the polarization gradient resonance is large enough to play a dominant role in the cooling process, resulting in a regular one-temperature Maxwellian distribution of atoms [Fig. \ref{Cloud_shape} (d)].
For much smaller saturation parameters the number of trapped atoms  rapidly decreases \cite{Tm_MOT}, making observation of
other MOT regimes difficult.

Double structures were previously reported for Rb \cite{Rb_double-structure_potential}, Dy \cite{Dy_double_structure} and Ca \cite{Hansen2003} MOTs. The specific cloud density profile in the Rb MOT  was caused by a bi-harmonical confining potential coming from the sub-Doppler part of a magneto-optical trapping force \cite{Rb_double-structure_potential}.  For the Dy and Ca MOTs the  double structure comes directly from the two-component Maxwellian momentum distribution, which resembles our case. The Dy MOT operates at the broad transition ($\Gamma = 2 \pi \times 32$\,MHz) and the two-temperature distribution appears only due to the influence of the magnetic field on the cooling process. The Ca MOT works at the narrow transition ($\Gamma = 2 \pi \times 57$\,kHz) involving atoms in a meta-stable state. As a result the atomic cloud doesn't attain thermal equilibrium, which results in a non-Maxwellian momentum distribution.
The double-structured MOT of the same origin as in our case was predicted for Mg atoms cooled on the $3^3P_2 \to 3^2D_3$ transition \cite{prudnikov2016lp,prudnikov2016}.

\section{Semiclassical and quantum descriptions of the double-structure regime}
\label{simulations}

\subsection{Semiclassical approach, molasses}

\begin{figure}[!t]
\center{\resizebox{0.45\textwidth}{!}{\includegraphics{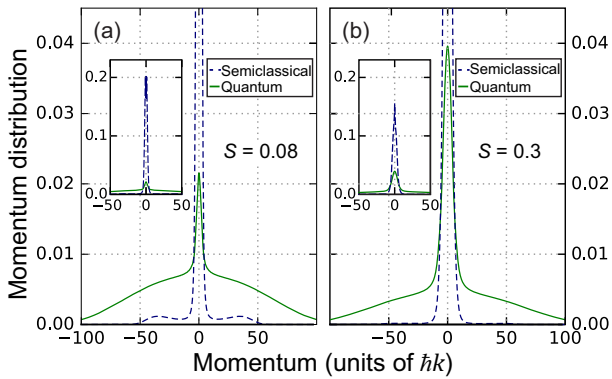}}}
\caption{Stationary momentum distribution of Tm atoms in the
one-dimensional $\sigma^+-\sigma^-$ optical molasses, showing a comparison of the
semiclassical calculations (blue dashed line) and full quantum treatment (green solid line).The
insets show full scale views.
Calculations are performed for (a) $S=0.08$ ($I = 5$\,mW$/$cm$^2$, $\Delta=-7\Gamma$) and (b) $S=0.3$ ($I = 20$\,mW$/$cm$^2$, $\Delta=-7\Gamma$).
}
\label{momentum_distribution}
\end{figure}

To analyze in detail the double-structure regime, we performed numerical simulations of the cooling process.
First, we tried to reproduce the observed momentum distribution using the semiclassical model of one-dimensional optical molasses in the $\sigma^+-\sigma^-$ configuration at zero magnetic field \cite{sub-Doppler_theory}.
The calculation was performed at the detuning $\Delta=-7\Gamma$ for intensities of $5$\,mW$/$cm$^2$ ($S_0=15.6$, $S=0.08$) and $20$ \,mW$/$cm$^2$ ($S_0=62.5$, $S=0.3$).
Results of the calculation are shown in Fig.\,\ref{momentum_distribution} with dashed lines.

As expected, the atomic momentum distribution consists of a narrow peak around zero momentum representing a cold atomic fraction and  a wide pedestal corresponding to a hot fraction of atoms.
We find that momentum profile of the hot fraction strongly deviates from a Gaussian [Fig.\,\ref{momentum_distribution}(a)].
Also, the relative number of hot atoms (pedestal area) rapidly decreases for the higher saturation parameter and is below $5\%$ at $S=0.3$ [Fig.\,\ref{momentum_distribution}(b)].
Both of these results contradict experimental observations which means that the semiclassical treatment is not valid in our case.

\begin{figure}[h!]
\resizebox{0.45\textwidth}{!}{
\includegraphics{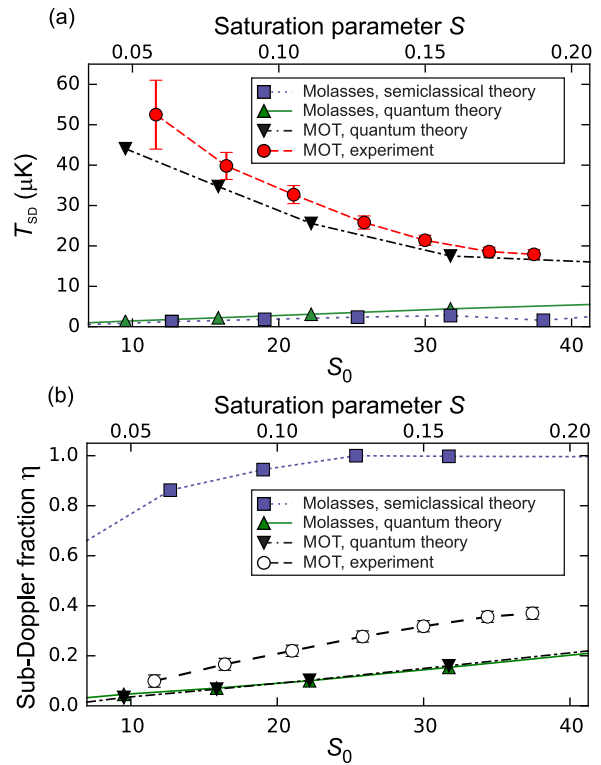}
} \caption {(a) Temperature of the sub-Doppler cooled atoms $T_{\rm SD}$ and (b) the sub-Doppler fraction $\eta_{\rm cold}$ depending on the cooling light intensity  $I$ ($S_0 \propto  I$). We show results of the semiclassical approach for molasses (blue squares), the quantum
approach for molasses (green upward triangles) and the quantum approach with the MOT magnetic field taken into account  (black downward triangles). Experimental results for $T_{\rm SD}$ (red circles) and $\eta_{\rm cold}$ (white circles)  are taken from Fig.\,\ref{Experiment}. The magnetic field gradient equals $b=0.07$\,T$/$m, the detuning is $\Delta = -7\Gamma$.
}
\label{TvsI}
\end{figure}

\subsection{Quantum approach, molasses}
As an alternative approach, we used the full quantum treatment described previously in Refs. \cite{Prudnikov2007,Prudnikov2011}.
In contrast to the  semiclassical case, this approach takes fully into account the recoil effect.
We numerically solve the master equation for atomic density matrix which gives full information of the system.
As in the previous case, calculations are performed for the one-dimensional configuration in the absence of magnetic fields. Results of the simulations are shown in Fig.\,\ref{momentum_distribution} together with the semiclassical ones. One can see that the quantum treatment also predicts a significantly non-Maxwellian momentum distribution consisting of two distinct sub-ensembles. There is a cold fraction accumulated around zero momentum and a wide pedestal spreading up to $\pm 100\hbar k$.

Despite qualitative similarity, the results of the quantum and the semiclassical approaches strongly differ mainly by the shape and the weight of the pedestal.
The momentum distribution obtained by the quantum approach can be well approximated by two independent Gaussian functions, one for the central peak (cold atomic fraction) and one for the pedestal (hot fraction). It allows us to deduce individual temperatures of two sub-ensembles $T_{\rm SD}$ and $T_{\rm D}$ as well as the fraction $\eta$ of the sub-Doppler cooled  atoms, similar to the experimental case (Section \ref{regimes}).

A comparison of the two theoretical approaches (semiclassical and quantum) with the experiment is presented in Fig.\,\ref{TvsI}. The results of quantum theory (``molasses, quantum theory'') and semiclassical theory (``molasses, semiclassical theory'') give similar prediction for the temperature of the sub-Doppler cooled atoms $T_{\rm SD}$ in the range of 0.5-5\,$\mu$K, which grows with the saturation parameter $S$. Both theories predict much lower temperatures (1-5\,$\mu K$) compared to what have been observed  in the experiment (20-50\,$\mu K$).

\subsection{Quantum approach, MOT}

Along with the recoil effect, one has to account for the  magnetic field of  the MOT.  Indeed,  the magnetic field can be neglected only if the Zeeman splitting of the ground state  $\Delta E/\hbar$ is much smaller compared to  the spectral width of the sub-Doppler resonance $\Gamma S$ \cite{sub-Doppler+MF}:
\begin{equation}
    \Delta E/\hbar = \mu_Bg_FB/\hbar\ll \Gamma S\,,
    \label{Zeeman_splitting}
\end{equation}
where $\mu_B$ is the Bohr magneton, $g_F$ is the Land\'e $g$-factor of the ground state and $B$ is the magnetic field.
In our experiment, $S \sim 0.1$ and the magnetic field gradient equals $b = 0.07$\,T$/$m, so the magnetic field can be neglected  only in a small area of 10\,$\mu$m  around the trap center. A typical cloud size of the Tm MOT is about 100\,$\mu$m which means that the magnetic field significantly impacts the cooling process by decreasing the polarization-gradient cooling efficiency.

To take into account the influence of the magnetic field we applied the density matrix method previously developed in
\cite{Prudnikov2007,Prudnikov2011} with an assumption that the
motion of the atoms in the MOT is much slower compared to the cooling rate. Indeed, for our range of parameters
 the MOT
oscillation frequency is close to 1\,kHz, while the recoil frequency, describing the cooling rate ($\tau_{\rm cool} \sim
 \omega_{\rm rec}^{-1}$) is $\omega_{\rm rec} /2\pi= \hbar k^2/4 \pi m = 4.2$ kHz. Thus, we can assume
adiabatic motion of the atoms in the magneto-optical potential with the
equilibrium momentum distribution determined by the local magnetic field
at the position $z$. This assumption allows us to derive the total momentum distribution of our atomic ensemble by averaging partial contributions from distinct sub-ensembles distributed along the $z$-axis.

Simulations show that the magnetic field significantly modifies the temperature and
spatial distribution of the cold sub-Doppler fraction of atoms.
The temperature $T_{\rm SD}$ calculated in the presence of the magnetic field is shown in Fig.\,\ref{TvsI}(a) as ``MOT, quantum theory''.  In the presence of the magnetic field the temperature grows up to one order of magnitude indicating that the polarization gradient cooling  becomes less efficient.  Also, the slope changes sign: now  $T_{\rm SD}$  decreases at higher intensities $I$. As one can see from Fig.\,\ref{TvsI}(a), results of the calculations in the presence of a magnetic field reproduce our experimental observations well.

The specific behavior of $T_{\rm SD} (S_0)$ can be qualitatively explained by the following. At higher intensities $I$, the width of the sub-Doppler resonance  becomes larger and the criterion Eq. (\ref{Zeeman_splitting}) becomes valid for a larger volume. As a result, the efficiency of the sub-Doppler cooling increases and the momentum distribution approaches the one predicted by the  molasses theory.

As shown in Fig.\,\ref{TvsI}(b), the fraction of the sub-Doppler cooled atoms $\eta$ predicted by the quantum theory is significantly smaller than the semiclassical one. Taking the magnetic field into account generally does not impact this fraction, so the results for the molasses and for the MOT nearly coincide. We see that the experimental results again are much better reproduced by the quantum approach rather than by the semiclassical one. The remaining discrepancy can be explained by the fact that simulations take into account the whole initial atomic ensemble, while in the experiment the hottest atoms cannot be trapped in the  finite MOT potential.

The latter also explains the significant discrepancy between the temperature $T_{\rm D}$ deduced from the simulations (typically about 1\,mK) and from the experiment (100\,$\mu$K). Indeed, the loss of the fastest atoms results in a significant change of temperature and makes the corresponding comparison uninformative.

Thus, using the quantum theory for a one-dimensional atomic gas in the presence of a magnetic field we adequately reproduced competitive processes between the Doppler and the sub-Doppler cooling mechanisms taking place in the second-stage Tm MOT operating at 530.7\,nm.

To the best of our knowledge, one-dimensional models of the MOT give the same result as three-dimensional models at the low saturation limit \cite{3D_MOT_Gajda, 3D_MOT_Minogin, 3D_MOT_Prudnikov}. Because of this, we restrict our calculations to the one-dimensional case for simplicity.

\section{Conclusion}
\label{conclusion}

We experimentally study the second-stage Tm MOT operating at the 530.7\,nm transition with an intermediate spectral linewidth of 350\,kHz. This gives an interesting opportunity to observe  three distinct MOT regimes depending on the saturation parameter $S$. The most intriguing regime can be observed for $S$ ranging from 0.05 to 0.5 when the cloud consists of the sub-Doppler and the Doppler cooled fractions. Both processes compete depending on the saturation parameter $S$ with the sub-Doppler cooled fraction increasing at higher $S$.

We expand previously reported one-dimensional full-quantum treatment by taking the MOT magnetic field into account to numerically calculate velocity distribution of atoms in the MOT.
We show that this model reproduces the experimental data well and can be successfully applied to describe kinetics of atoms in the MOT.

\section{Acknowledgments}

The authors acknowledge support from RFBR grants \#15-02-05324 and \#16-29-11723. The research work of O.N.P. was supported by RSF (project \#16-12-00054).

\bibliography{references}

\end{document}